\documentclass{article}

\usepackage{arxiv}

\usepackage[utf8]{inputenc} 
\usepackage[T1]{fontenc}    
\usepackage{url}            
\usepackage{booktabs}       
\usepackage{amsfonts}       
\usepackage{nicefrac}       
\usepackage{microtype}      
\usepackage{graphicx}
\usepackage{doi}
\usepackage{amsmath}

\title{Exploring the role of polarization in fiber-based quantum sources}

\author{ {\hspace{1mm}Carla M.~Brunner} \\
	Max Planck Institute for the Science of Light,\\
	Staudtstr. 2, 91058 Erlangen, Germany\\
    \\
    Friedrich-Alexander-University Erlangen-Nürnberg,\\
    Staudtstr. 7, 91058 Erlangen, Germany\\
    \\
	\texttt{carla.brunner@mpl.mpg.de} \\
	\And
	{\hspace{1mm}Nicolas Y.~Joly} \\
	Friedrich-Alexander-University Erlangen-Nürnberg,\\
    Staudtstr. 7, 91058 Erlangen, Germany\\
    \\
	Max Planck Institute for the Science of Light,\\
	Staudtstr. 2, 91058 Erlangen, Germany\\
    \\
	\texttt{nicolas.joly@mpl.mpg.de} \\
}

\date{October 17, 2024}

\renewcommand{\headeright}{}
\renewcommand{\undertitle}{}

\begin{document}
\maketitle

\begin{abstract}
	Optical fibers constitute an attractive platform for the realization of nonlinear and quantum optics processes. Here we show, through theoretical considerations, how polarization effects of both third-order parametric down-conversion and four-wave-mixing in optical fibers may be exploited to enhance detection schemes. We apply our general framework specifically to the case of tapered fibers for photon triplet generation, a long-standing goal within quantum optics, and obtain explicit expectation values for its efficiency. A quantitative investigation of four-wave-mixing in a microstructured solid-core fiber provides significant consequences for the role of polarization in experimental design.
\end{abstract}


\section{Introduction}
Over recent decades, we have been witness to not only consequential advancements in both the production and characterization of microstructured optical fibers \cite{Petrovich2005, Liu2023} but also in the area of quantum communication, which frequently builds on accessible sources of entangled photons \cite{Brambila2023}. Specialty fibers may be employed for the generation of photon pairs or even photon triplets through nonlinear interactions \cite{GarayPalmett2023, Shukhin2020}. Several qualities make them a compelling tool for these applications, including the tight confinement of the interacting fields and long interaction lengths \cite{McGarry2024}. Over the last decades, they have been developed into an incredibly versatile range of various shapes and forms, for example photonic-crystal fibers (PCF), gas- or liquid- filled hollow-core fibers (HCF), twisted fibers, as well as microstructured nano- or tapered fibers \cite{Birks1997, Knight1998, Markos2017, Grubsky2007}. They can serve as implementations of photon-pair generation or twin-beam generation \cite{Finger2015} through the process of four-wave-mixing (FWM) and have been proposed as a possible technique to enable photon triplet generation through third-order parametric down-conversion (TOPDC) \cite{Cavanna2020}.\\
Both FWM and TOPDC rely on the third-order nonlinearity of the interaction medium and require fulfillment of the respective phase matching condition. While building wavelength tunable fiber-based photon-pair sources has been achieved, triplet generation through TOPDC in the visible domain remains a long-standing goal within quantum optics.\\
The generation of nonclassical states of light, especially higher-order entanglement, is expected to have far-reaching implications for tests of fundamental quantum mechanics \cite{Banaszek1997}, quantum communication \cite{Gisin2007, Lo2014} and quantum computing \cite{Browne2005}. Three-photon state generation has been successfully implemented in the microwave domain \cite{Chang2020}. In the visible region, there have been advances in the direction of triplet generation using varying experimental setups, for example depending on cascaded down-conversion processes in nonlinear crystals \cite{Hbel2010, Hamel2014}, showing cubic optical parametric down-conversion in a crystal \cite{Douady2004}, building a triplet source on the basis of quantum dot molecules \cite{Khoshnegar2017}, designing integrated waveguides \cite{Moebius2016} or hybrid PCFs \cite{Cavanna2016} optimized for TOSPDC.\\
A promising platform for an implementation of direct TOSPDC is given by tapered optical fibers \cite{Corona2011,Cavanna2020}. The experimental proposal assumes a standard silica fiber that is tapered down to an outer diameter of just $790.36$ nm and pumped by a narrowband laser at $532$ nm. Intermodal phase matching is then expected between the higher-order HE$_{12}$ pump mode and the fundamental HE$_{11}$ signal mode at $1596$ nm. An adiabatic taper transition is required in order to keep the higher-order pump mode throughout the fiber waist. An experimental realization has to this date been prohibited by the extremely low conversion efficiency.\\
The strength of nonlinear interactions in optical fibers is commonly described by an effective area $A_{\text{eff}}$, related to the modal overlap of interacting fields \cite{Wadsworth2004} and can be derived from overlap integrals\cite{Stolen1982}. Typically, descriptions involve only a single linear polarization. However, to successfully implement systems that use optical fibers to give rise to nonlinear and quantum processes such as FWM and TOPDC, it is unquestionably valuable to be able to theoretically describe all important aspects, particularly including different polarization states. This can in turn lead to the ability to optimize experiments and detection settings. Here we present an approach that we have developed in the context of triplet generation in tapered optical fibers. We will show that it can similarly be used to explain effects that pertain to seeded TOPDC in tapered and FWM in other types of specialty fibers.\\
We first establish a quantization procedure specifically for the situation of optical fiber modes and apply these expressions to derive polarization-dependent rates of triplet generation. Subsequently, we discuss consequences for optimized detection systems and seeded triplet generation. We conclude by carrying out a quantitative analysis of polarization effects in the FWM case, coinciding with prior studies of classical models of pulse propagation and supercontinuum generation in optical fibers based on the nonlinear Schrödinger equation \cite{Poletti2008}.

\section{Quantization of the electromagnetic field in optical fibers}
The derivation of a theoretical triplet rate in optical fibers necessitates expressions for quantized optical fiber modes involved in the nonlinear interaction. To that end, we start off with Maxwell's equations and follow a similar quantization procedure for fields in inhomegeneous dielectrics as presented by Milonni \cite{Milonni2019}. Using the gauge fixing $\Phi = 0,~\nabla\cdot[\epsilon(\textbf{r},\omega)\textbf{A}] = 0$ for the electric scalar potential $\Phi$ and the magnetic vector potential $\textbf{A}$, where $\epsilon$ is the generally position- and frequency-dependent relative permittivity, a generalization of the Coulomb gauge, the classical vector potential is a solution to the equation of motion 
\begin{equation}
    \frac{\epsilon(\textbf{r},\omega)}{c^2}\frac{\partial^2\textbf{A}}{\partial t^2} + \nabla\times(\nabla\times\textbf{A}) = 0,
\end{equation}
with $t,c$ denoting time and vacuum speed of light, and can be expanded in a set of spatial vectorial eigenmodes $\mathbf{f}_{\xi}(\mathbf{r})$,
\begin{equation}
    \textbf{A}(\textbf{r},t) = \sum_{\xi} \alpha_{\xi}(t)\textbf{f}_{\xi}(\textbf{r}) + \alpha^*_{\xi}(t)\textbf{f}^*_{\xi}(\textbf{r}),
\end{equation}
determined by the solutions of the classical Maxwell equations, where the expansion coefficients $\alpha_{\xi}(t) = \alpha_{\xi}(t_0)e^{-i\omega_{\xi}(t-t_0)},
    \alpha^*_{\xi}(t) = \alpha^*_{\xi}(t_0)e^{i\omega_{\xi}(t-t_0)}$, for some initial time $t_0$, oscillate sinusoidally in time and the subscript $\xi$ runs over the possible solutions of
\begin{equation}\label{eq:eigenmode_eq}
    -\frac{\epsilon(\textbf{r})\omega_{\xi}^2}{c^2}\textbf{f}_{\xi} + \nabla\times(\nabla\times\textbf{f}_{\xi}) = 0.
\end{equation}
Since $\Phi=0$, the expansion gives rise to the physical electric and magnetic fields as
\begin{equation}
\begin{split}
    \textbf{E}(\textbf{r},t) &= -\nabla\Phi-\frac{\partial\textbf{A}}{\partial t} = \sum_{\xi} i\omega_{\xi}\alpha_{\xi}(t)\textbf{f}_{\xi}(\textbf{r}) - i\omega_{\xi} \alpha^*_{\xi}(t)\textbf{f}^*_{\xi}(\textbf{r}),\\
    \textbf{B}(\textbf{r},t) &= \nabla\times\textbf{A} = \sum_{\xi} \alpha_{\xi}(t)(\nabla\times\textbf{f}_{\xi}(\textbf{r})) + \alpha^*_{\xi}(t)(\nabla\times\textbf{f}^*_{\xi}(\textbf{r})).
\end{split}
\end{equation}
Let us now consider a single mode, $\textbf{A}=\alpha\textbf{f}+\alpha^*\textbf{f}^*$ . The classical field energy $\mathcal{E}$ is dependent on the material dispersion properties (for a derivation of this expression see suppl. material (SM), eq. S8) reads
\begin{equation}
    \mathcal{E} = 2\omega^2|\alpha^2|\int_V d^3\textbf{r}\left[\frac{\frac{d}{d\omega}(\epsilon\omega)+\epsilon}{2}\epsilon_0|\textbf{f}|^2\right]
\end{equation}
and we choose the normalization
\begin{equation}
    \int_V d^3\textbf{r}\left[\frac{\frac{d}{d\omega}(\epsilon\omega)+\epsilon}{2}\epsilon_0|\textbf{f}|^2\right] = 1.
\end{equation}
The electromagnetic field energy is equated to the Hamiltonian of a quantum Harmonic Oscillator, $H_{HO} = \frac{1}{2}(p^2+\omega^2q^2)$, by writing the canonically conjugate position and momentum variables as $q(t) = i(\alpha(t)-\alpha^*(t)),p(t) = \omega(\alpha(t)+\alpha^*(t))$, obeying the Hamiltonian equations of motion. We now replace the classical variables by their quantum counterparts, $\alpha(t) \rightarrow C\hat{a}(t)$, $\alpha^*(t) \rightarrow C^*\hat{a}^{\dagger}(t)$, with the canonical commutation relations $[\hat{a}_{\xi},\hat{a}_{\xi'}^{\dagger}]=\delta_{\xi\xi'}$ . From $[q,p]=i\hbar$ it follows that the constant $C_{\xi}=\sqrt{\frac{\hbar}{2\omega_{\xi}}}$. Hence,
\begin{equation}
\begin{split}
    \hat{\textbf{A}}(\textbf{r},t) &= \sum_{\xi} \sqrt{\frac{\hbar}{2\omega_{\xi}}}\left[\hat{a}_{\xi}(t)\textbf{f}_{\xi}(\textbf{r})+\hat{a}_{\xi}^{\dagger}(t)\textbf{f}^*_{\xi}(\textbf{r})\right],\\
    \hat{\textbf{E}}(\textbf{r},t) &= i \sum_{\xi} \sqrt{\frac{\hbar\omega_{\xi}}{2}}\left[\hat{a}_{\xi}(t)\textbf{f}_{\xi}(\textbf{r})-\hat{a}_{\xi}^{\dagger}(t)\textbf{f}^*_{\xi}(\textbf{r})\right],
\end{split}
\end{equation}
where the first terms correspond to the positive and the second terms to the negative frequency components respectively. In the present case of optical fibers, the eigenmodes $\textbf{f}$ are related to the mode field components as will be detailed below.
\section{The rate of triplet generation}\label{sec:rate_triplet_generation}
Adopting the expressions or electric field operators in optical fibers, in this section, we derive the triplet generation rate in fibers. To this end, we assume low conversion efficiency so that the rate of triplet rate per signal modes configuration is given by Fermi's Golden Rule \cite{Okoth2019},
\begin{equation}
    \Gamma_{\text{triplet}} = \frac{2\pi}{\hbar^2}~\delta(\Delta\omega)~|\langle \alpha(N_p-1),1,1,1|\hat{H}_I|\alpha(N_p),0,0,0\rangle|^2,
\end{equation}
where $N_p$ is the mean pump photon number and $\hat{H}_I$ is the interaction Hamiltonian,
\begin{equation}\label{eq:interaction_hamiltonian}
    \hat{H}_I=-\epsilon_0\cdot4\cdot3!\cdot\chi^{(3)}_{ijkl}\int_{V_I}d^3\textbf{r}~\hat{E}^{(+)}_{p,i}\hat{E}^{(-)}_{s_1,j}\hat{E}^{(-)}_{s_2,k}\hat{E}^{(-)}_{s_3,l}~+ \text{h.c.}.
\end{equation}
$\chi^{(3)}$ is the third-order nonlinear susceptibility tensor. Here we assume only a weak dependency on the signal frequencies, i.e. $\chi^{(3)}\equiv\chi^{(3)}(\omega_p; \omega_s, \omega_s, \omega_s)$, and we use Einstein's summation convention to sum over the indices $i,j,k,l=x,y,z$ or $i,j,k,l=r,\phi,z$ in cylindrical coordinates. $V_I$ denotes the interaction volume of a fiber of length $L$, e.g. the fused silica core in case of a tapered fiber. The initial and final states are $\psi_i=|\alpha(N_p),0,0,0\rangle$ and $\psi_f=|\alpha(N_p-1),1,1,1\rangle$, where $|\alpha(N)\rangle$ is the coherent pump state with a mean photon number $N$.
The total rate of triplet generation is obtained by summation over all signal mode configurations of the degrees of freedom, including polarization,
\begin{equation}
    R_{\text{triplet}} = \sum_{\xi_{s_1}}\sum_{\xi_{s_2}}\sum_{\xi_{s_3}} \frac{2\pi}{\hbar^2}~\delta(\Delta\omega)~|\langle \hat{H}_I\rangle|^2.
\end{equation}
Considering the quantized electric field of just a single fiber mode and polarization, a one-to-one relationship between the labels $\xi$ and propagation constants $\beta$ can be established. The summation is hence written as $\sum_{\xi}\rightarrow\sum_{\sigma}\frac{L}{2\pi}\int d\beta$, since each propagation constant $\beta$ covers a phase-space volume of $\frac{2\pi}{L}$. Here, $\sigma$ denotes one of two orthogonal polarization states. Therefore,
\begin{equation}
\begin{split}
    R_{\text{triplet}} & = \sum_{\sigma_{s_1},\sigma_{s_2},\sigma_{s_3}}\left(\frac{L}{2\pi}\right)^3\iiint d \beta_{s_1}d \beta_{s_2}d \beta_{s_3}\frac{2\pi}{\hbar^2}~\delta(\Delta\omega)~|\langle \hat{H}_I\rangle|^2\\
    &= \sum_{\sigma_{s_1},\sigma_{s_2},\sigma_{s_3}}\left(\frac{L}{2\pi}\right)^3\iiint d \omega_{s_1}d \omega_{s_2}d \omega_{s_3}\left.\frac{d\beta}{d\omega}\right\vert_{\omega_{s_1}}\left.\frac{d\beta}{d\omega}\right\vert_{\omega_{s_2}}\left.\frac{d\beta}{d\omega}\right\vert_{\omega_{s_3}}\frac{2\pi}{\hbar^2}~\delta(\Delta\omega)~|\langle \hat{H}_I\rangle|^2\\
    &= \sum_{\sigma_{s_1},\sigma_{s_2},\sigma_{s_3}}\left(\frac{L}{2\pi}\right)^3\iint d \omega_{s_1}d \omega_{s_2}\left.\left[\left.\frac{d\beta}{d\omega}\right\vert_{\omega_{s_1}}\left.\frac{d\beta}{d\omega}\right\vert_{\omega_{s_2}}\left.\frac{d\beta}{d\omega}\right\vert_{\omega_{s_3}}\frac{2\pi}{\hbar^2}~|\langle \hat{H}_I\rangle|^2\right]\right\vert_{\Delta\omega=0}
\end{split}
\end{equation}
and a finite detection bandwidth can be taken into account by adjusting the integration limits accordingly. Separating the integrals in the expression for $\hat{H}_I$ (eq. \ref{eq:interaction_hamiltonian}) into longitudinal integration over the fiber length $L$ and transversal integration over the fiber cross-section $A_I$ (see SM, eq. S9), $\hat{H}_I$ evaluates to
\begin{equation}
\begin{split}
    \hat{H}_I &= 24\chi^{(3)}_{ijkl}\epsilon_0~\sqrt{\frac{\hbar\omega_p}{2}}\sqrt{\frac{\hbar\omega_{s_1}}{2}}\sqrt{\frac{\hbar\omega_{s_2}}{2}}\sqrt{\frac{\hbar\omega_{s_3}}{2}}~a_pa^{\dagger}_{s_1}a^{\dagger}_{s_2}a^{\dagger}_{s_3}~\frac{1}{L^2}\frac{1}{\sqrt{M_pM_{s_1}M_{s_2}M_{s_3}}}\\
    &~~~~~~~~~\times\int_{L}dze^{i(\beta_p-\beta_{s_1}-\beta_{s_2}-\beta_{s_3})z}~\int_{A_I}dxdy~\mathcal{I}_{ijkl}(x,y) + \text{h.c.}.\\
\end{split}
\end{equation}
Here we have used the solutions $\textbf{e}(x,y)$ for electromagnetic fields of modes propagating in cylindrical step-index fibers as given in \cite{Snyder1984}, for which $\textbf{e}(x,y)e^{i\beta z}$ are solutions to the wave equation (eq. \ref{eq:eigenmode_eq}) and introduced
\begin{equation}\label{eq:normalization_def}
\begin{split}
    \textbf{f} =& \frac{1}{\sqrt{LM}}e^{i\beta z}~\textbf{e},\\
    M =& \int_{A_{\infty}}~dxdy~\frac{\frac{d}{d\omega}(\epsilon\omega)+\epsilon}{2}\epsilon_0~\textbf{e}(x,y)\cdot\textbf{e}^*(x,y),\\
    1 =& \int_V d^3\textbf{r}\left[\frac{\frac{d}{d\omega}(\epsilon\omega)+\epsilon}{2}\epsilon_0|\textbf{f}|^2\right]\\
    \mathcal{I}_{ijkl} (x,y) :=&~ e_{p,i}(x,y)e_{s_1,j}^*(x,y)e_{s_2,k}^*(x,y)e_{s_3,l}^*(x,y).
\end{split}
\end{equation}
Continuing with the evaluation of the triplet rate, we have
\begin{equation}
    \langle\alpha(N_p-1)|a_p|\alpha(N_p)\rangle \approx\sqrt{N_p}
\end{equation}
in the limit of large pump photon numbers $N_p$ (see SM, eq. S10) and $\langle1,1,1|\hat{a}^{\dagger}_{s_1}\hat{a}^{\dagger}_{s_2}\hat{a}^{\dagger}_{s_3}|0,0,0\rangle=1$, so that
\begin{equation}
\begin{split}
    |\langle\hat{H}_I\rangle|^2 &= 24^2N_p\epsilon_0^2\frac{\hbar^4\omega_p\omega_{s_1}\omega_{s_2}\omega_{s_3}}{2^4}\frac{1}{L^4}\frac{1}{M_pM_{s_1}M_{s_2}M_{s_3}}\\
    &~~~~~~~~~\times\left|\int_{L}dze^{i(\beta_p-\beta_{s_1}-\beta_{s_2}-\beta_{s_3})z}\right|^2~\left|\chi^{(3)}_{ijkl}\int_{A_I}dxdy~\mathcal{I}_{ijkl}(x,y)\right|^2.\\
\end{split}
\end{equation}
Furthermore, we define the mode overlap as
\begin{equation}
    \mathcal{O} = \left|\hat{\chi}^{(3)}_{ijkl}\int_{A_I}dxdy~\mathcal{I}_{ijkl}(x,y)\right|^2,
\end{equation}
where $\hat{\chi}^{(3)}_{ijkl}$ is the normalized third-order nonlinear susceptibility,
$\hat{\chi}^{(3)}_{ijkl}:=3\chi^{(3)}_{ijkl}/\chi^{(3)}_{xxxx}$. The phase mismatch factor with wavevector mismatch $\Delta\beta:=\beta_p-\beta_{s_1}-\beta_{s_2}-\beta_{s_3}$ evaluates to
\begin{equation}
    \left\vert\int_{L} dz~e^{-i(\beta_p-\beta_{s_1}-\beta_{s_2}-\beta_{s_3})z}\right\vert^2 = L^2\text{sinc}^2\left(\frac{\Delta\beta L}{2}\right).
\end{equation}
Thus,
\begin{equation}
    |\langle\hat{H}_I\rangle|^2 = \frac{36N_p(\chi_{xxxx}^{(3)}/3)^2\epsilon_0^2\hbar^4\omega_p\omega_{s_1}\omega_{s_2}\omega_{s_3}}{L^2M_pM_{s_1}M_{s_2}M_{s_3}}~\text{sinc}^2\left(\frac{\Delta\beta L}{2}\right)~\mathcal{O}.\\
\end{equation}
Finally, the triplet generation rate becomes
\begin{equation}\label{eq:triplet_rate}
\begin{split}
    R_{\text{triplet}} &= \sum_{\sigma_{s_1},\sigma_{s_2},\sigma_{s_3}}\frac{P_pL^2n_p(\chi_{xxxx}^{(3)})^2\epsilon_0^2\hbar}{\pi^2c}\\
    &~~~~~~\times\iint d \omega_{s_1}d \omega_{s_2}\left.\left[\frac{\omega_{s_1}\omega_{s_2}\omega_{s_3}}{M_pM_{s_1}M_{s_2}M_{s_3}}\left.\frac{d\beta}{d\omega}\right\vert_{\omega_{s_1}}\left.\frac{d\beta}{d\omega}\right\vert_{\omega_{s_2}}\left.\frac{d\beta}{d\omega}\right\vert_{\omega_{s_3}}\text{sinc}^2{\left(\frac{\Delta\beta L}{2}\right)}\mathcal{O}\right]\right\vert_{\Delta\omega=0}.
\end{split}
\end{equation}
Using this result, we are now able to calculate the expected triplet rate for an ideal fiber with the degenerate phase-matching radius of 395.18 nm, a taper waist length of 4 cm and a pump power of 100 mW, assuming $\chi_{xxxx}^{(3)}=2.8\times10^{-22}$m$^2$/V$^2$. Independent of the pump polarization state, we obtain a total triplet rate of 5.1 Hz. Excitation of the pump mode can be achieved through a number of methods, e.g. by conversion of the fundamental mode via long-period-gratings \cite{Zhao2016}. We can also draw a conclusion on the polarization state of the generated triplets, as will be discussed in the following section.\\
\subsection{Mode overlap and the role of polarization}\label{sec:mode_overlap}
\begin{figure}[htpb]
    \centering\includegraphics[width=\textwidth]{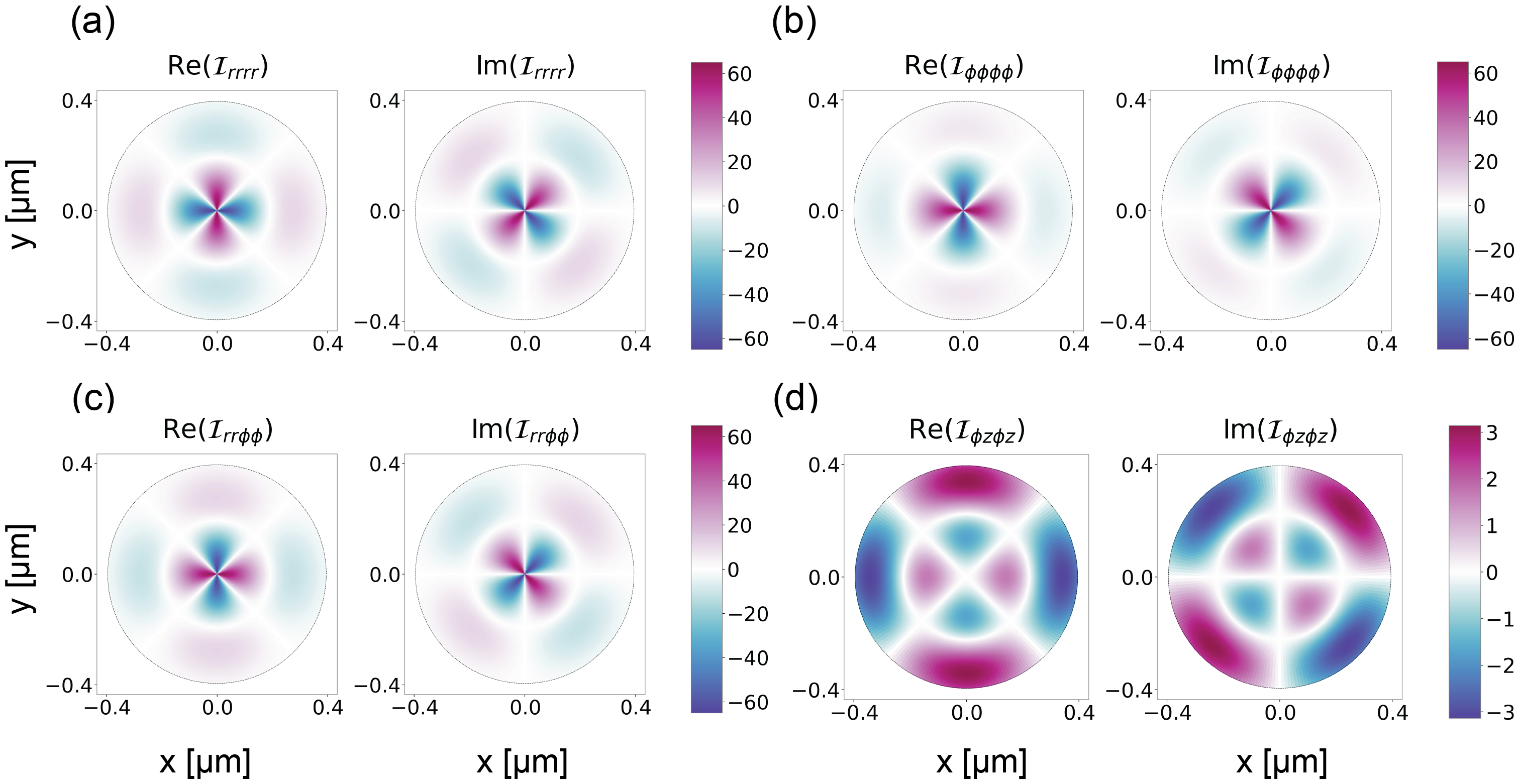}
    \caption{Real and imaginary parts of selected components of the mode overlap integrand $\mathcal{I}_{ijkl}$ for the L$\rightarrow$L,L,L configuration. The colorbars indicate the magnitudes of $\text{Re}[\mathcal{I}(x,y)]$ and $\text{Im}[\mathcal{I}(x,y)]$ respectively. Integrating the $\mathcal{I}_{ijkl}$ over the interaction area and summation gives a zero mode overlap and thus a zero triplet rate. \mbox{(a) $\mathcal{I}_{rrrr}=e_{p,r}e^*_{s_1,r}e^*_{s_2,r}e^*_{s_3,r}$,} \mbox{(b) $\mathcal{I}_{\phi\phi\phi\phi}=e_{p,\phi}e^*_{s_1,\phi}e^*_{s_2,\phi}e^*_{s_3,\phi}$,} \mbox{(c) $\mathcal{I}_{rr\phi\phi}=e_{p,r}e^*_{s_1,r}e^*_{s_2,\phi}e^*_{s_3,\phi}$,} \mbox{(d) $\mathcal{I}_{\phi z\phi z}=e_{p,\phi}e^*_{s_1,z}e^*_{s_2,\phi}e^*_{s_3,z}$.}}
    \label{fig:mode_overlap_triplet_subfigures_LLLL}
\end{figure}

\begin{figure}[htpb]
    \centering\includegraphics[width=\textwidth]{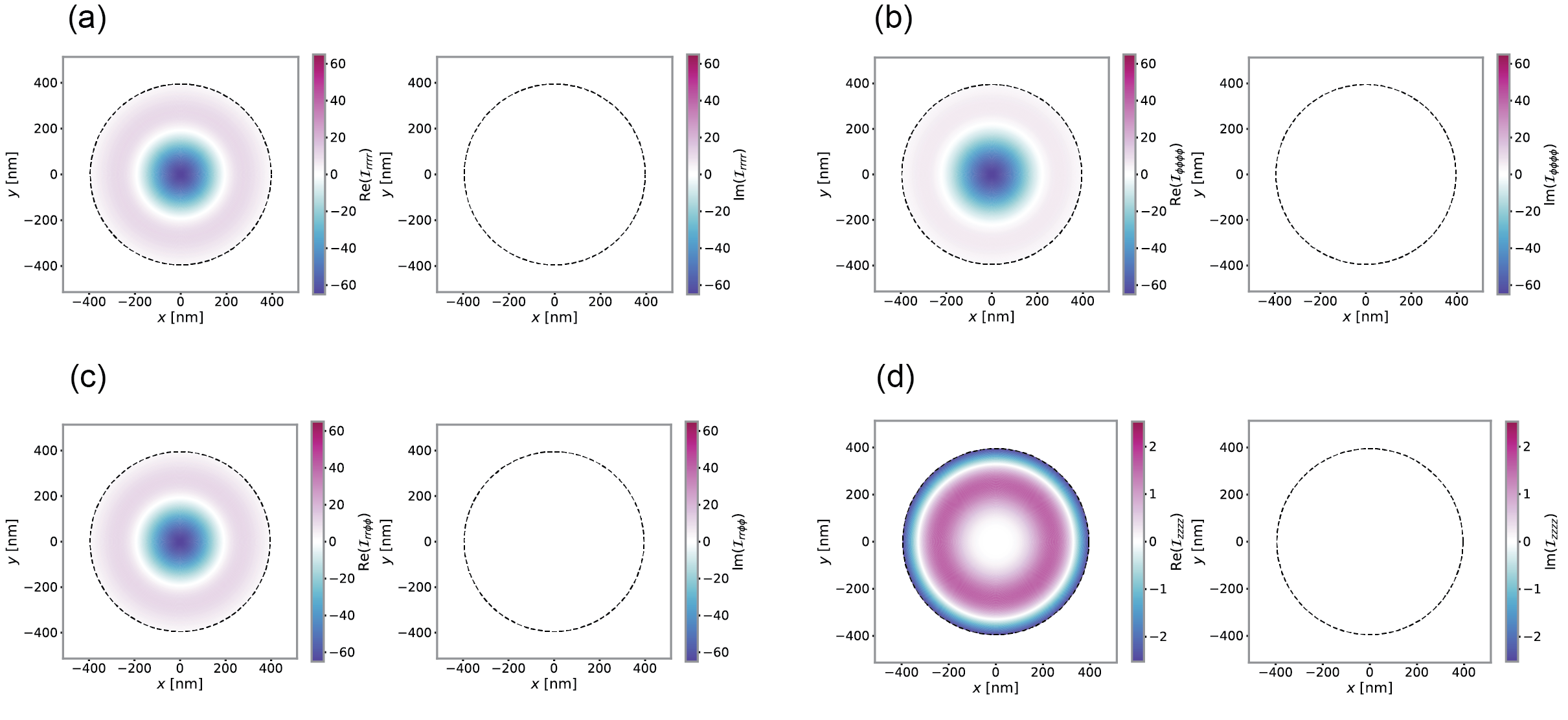}
    \caption{Real and imaginary parts of selected components of the mode overlap integrand $\mathcal{I}_{ijkl}$ for the L $\rightarrow$ L,L,R configuration. The colorbars indicate the magnitudes of $\text{Re}[\mathcal{I}(x,y)]$ and $\text{Im}[\mathcal{I}(x,y)]$ respectively. This configuration obeys angular momentum conservation. The resulting mode overlap and thus triplet rate are nonzero. \mbox{(a) $\mathcal{I}_{rrrr}=e_{p,r}e^*_{s_1,r}e^*_{s_2,r}e^*_{s_3,r}$,} \mbox{(b) $\mathcal{I}_{\phi\phi\phi\phi}=e_{p,\phi}e^*_{s_1,\phi}e^*_{s_2,\phi}e^*_{s_3,\phi}$,} \mbox{(c) $\mathcal{I}_{rr\phi\phi}=e_{p,r}e^*_{s_1,r}e^*_{s_2,\phi}e^*_{s_3,\phi}$,} \mbox{(d) $\mathcal{I}_{zzzz}=e_{p,z}e^*_{s_1,z}e^*_{s_2,z}e^*_{s_3,z}$.}}
    \label{fig:mode_overlap_triplet_subfigures_LLLR}
\end{figure}

While the fiber modes are not plane waves and can therefore not be ascribed a pure polarization state, the even and odd fiber modes are primarily polarized in the two linear orthogonal $x$,$y$ direction. In the following we will refer to these fiber modes as linearly polarized ('$x$','$y$') and their complex superpositions as circularly polarized ('L','R').
We first consider a linear pump polarization (e.g. $x$-polarization) and restrict the sum in eq. \ref{eq:triplet_rate} to different signal polarization configurations in the linear basis. We find that $75\%$ of triplets are generated in the $x$-$x$-$x$ polarization configuration (configuration A), $25\%$ in the indistinguishable $x$-$y$-$y$/$y$-$x$-$y$/$y$-$y$-$x$ configurations (configuration B), and none in the other configurations. In the case of a circularly polarized pump (e.g. L) and in the circular polarization basis, $100\%$ of triplets are generated in a configuration where one of the signal photons is right circularly polarized (R) and two of the signal photons are left circularly polarized (L) (configuration C). Note that this is a form of angular momentum conservation in the cylindrically symmetric tapered fiber as can be conjectured by Noether's theorem.\\
To further understand the emergence of angular momentum conservation in our expression, we look into the derived  triplet rate in more detail, specifically into the mode overlap $\mathcal{O}$ and the mode overlap integrand $\mathcal{I}_{ijkl} = e_{p,i}e_{s_1,j}^*e_{s_2,k}^*e_{s_3,l}^*$. Firstly, we consider pump and signal photons in the circular polarization basis and assume an L pump. In principle, one possible triplet polarization configuration would be L,L,L. In figure \ref{fig:mode_overlap_triplet_subfigures_LLLL}, we show several components of the mode overlap integrand with indices in cylindrical coordinates. 
Evidently, each of these components obeys a symmetry between positive and negative values for both the real and imaginary part so that integration over the mode overlap integrand components results in $0$. This is similarly true for the other components (see SM, Fig. S1, Fig. S2). Therefore, in the L$\rightarrow$L,L,L) configuration, the mode overlap and the triplet rate are $0$. The same holds for the L$\rightarrow$L,R,R (and permutations of the right-hand-side, Fig. S3) and L$\rightarrow$R,R,R configuration (Fig. S4), examples for which are given in the SM - but remarkedly not for the L$\rightarrow$L,L,R configuration, obeying angular momentum conservation (cf. Fig. \ref{fig:mode_overlap_triplet_subfigures_LLLR}).
\subsection{Optimization of the detection scheme}
To detect triplets, three superconducting nanowire single-photon detectors can be used to measure photon coincidence rates. The optical detection bandwidth is determined by choosing appropriate chromatic filters in the detection path which also suppress the pump.\\
Generated photon triplets will lead to identical arrival times at the detectors and thus to a coincidence-rate signal. However, the three triplet photons are not necessarily directed to different detectors. Furthermore, the detectors have a finite timing uncertainty (on the order of $50$ ps). We use a coincidence bin of approximately $\tau=$ $200$ ps. Additionally, they have a quantum efficiency $\eta_q$ which is close to but smaller than $1$.\\
The main contribution to accidental coincidences is fluorescence; the number of fluorescence photons is proportional to the pump power and fluorescence is emitted into random directions and hence into fiber modes of random polarization. We aim for the highest coincidence-to-accidental ratio (CAR) \cite{Davano2012}, given by 
\begin{equation}
    \text{CAR} = \frac{\text{triplet coincidence detection rate}}{\text{accidental coincidence detection rate}}.
\end{equation}
The triplet coincidence detection rate is proportional to the triplet rate $R_{\text{triplet}}$ but depends on the probabilistic splitting of the signal photons for a specific detection scheme, taking the quantum optical properties of dielectric beamsplitter into account. The accidental coincidence rate due to fluorescence is
\begin{equation}
    R_{\text{acc}} = \frac{1}{\tau} (R_{\text{fl}}^{D_1} \tau\eta_q) (R_{\text{fl}}^{D_2} \tau\eta_q) (R_{\text{fl}}^{D_3} \tau\eta_q) = \tau^2\eta_q^3R_{\text{fl}}^{D_1}R_{\text{fl}}^{D_2}R_{\text{fl}}^{D_3},
\end{equation}
where $R_{\text{fl}}^{D_i}$ denotes the effective fluorescence photon rates at detector $i$.\\
\subsubsection{Linear pump polarization}\label{subsec:optimization_detection_scheme_linear_pol}
The first detection scheme we want to consider is one where the pump is linearly ($x$-) polarized and only the $x$- polarized triplet photons are detected (Fig. \ref{fig:detection_schemes} (a): configuration L).
\begin{figure}[htpb]
    \centering\includegraphics[width=.7\textwidth]{detection_schemes3.png}
    \caption{Possible detection schemes for concidence detection of photon triplets. \mbox{(a) Linear polarization configuration L (x$\rightarrow$x,x,x) detection.} The pump at $532$ nm is linearly polarized. The signal IR photons emerging from the fiber, comprising triplet photons and fluorescence (noise) are directed onto a PBS so that only x-polarized photons remain in the detection path. Two additional BS split the triplet photons so that they can be collected by three single-photon detectors $D_1,D_2,D_3$. \mbox{(b) Circular polarization configuration C (L$\rightarrow$L,L,R) detection.} A circularly polarized beam serves as pump. The circularly polarized signal IR photons are first transformed to linear polarization states via a QWP. A subsequent PBS separates the previously L-polarized from the R-polarized photons. A BS separates the L-polarized photons so that they impinge on different detectors $D_2,D_3$.}
    \label{fig:detection_schemes}
\end{figure}
For this purpose, we direct the IR signal onto a polarizing beamsplitter (PBS) to filter out $y$- polarized photons. Subsequently, we then add a dielectric beamsplitter (BS) and a $50-50$ BS in one of the detection arms. To determine the optimal BS splitting ration, we need to consider that the general quantum optical description of the state transformation via a lossless BS is given by a unitary operator $U_{BS}(\theta)$, $|\text{out}\rangle=U_{BS}(\theta)~|\text{in}\rangle$ \cite{Mandel1995}. The operator acts on an incoming three-photon state as
\begin{equation}
\begin{split}
    U_{BS}(\theta)~|3,0\rangle_{\text{in}} &= U_{BS}(\theta)\frac{1}{\sqrt{6}}\left(\hat{a}_1^{\dagger}\right)^3|0,0\rangle_{\text{in}}\\
    &= \frac{1}{\sqrt{6}}\left(\hat{b}_1^{\dagger}\cos{\theta}+i\hat{b}_2^{\dagger}\sin{\theta}\right)^3|0,0\rangle_{\text{out}}\\
    &= \frac{1}{\sqrt{6}} \left[\cos^3\theta\left(\hat{b}_1^{\dagger}\right)^3 + 3i\cos^2\theta\sin\theta\left(\hat{b}_1^{\dagger}\right)^2\hat{b}_2^{\dagger}\right. \\
    &~~~~~~~~~~~~~~~~~~~~~\left.-3\cos\theta\sin^2\theta\hat{b}_1^{\dagger}\left(\hat{b}_2^{\dagger}\right)^2-i\sin^3\theta\left(\hat{b}_2^{\dagger}\right)^3\right]~~|0,0\rangle_{\text{out}}\\
    &= \cos^3\theta~|3,0\rangle_{\text{out}} + \sqrt{3}i\cos^2\theta\sin\theta~|2,1\rangle_{\text{out}} \\
    &~~~~~~~~~~~~~~~~~~~~~- \sqrt{3}\cos\theta\sin^2\theta~|1,2\rangle_{\text{out}}- i \sin^3\theta~|0,3\rangle_{\text{out}},
\end{split}
\end{equation}
where $\hat{a}_{1,2}^{\dagger}, \hat{b}_{1,2}^{\dagger}$ are the photon creation operators of indistuingishable photons of the same polarization at the two inputs and two outputs respectively. A necessary and sufficient condition for triplet coincidence detection is now a two-photon state in the reflection arm, since this two-photon state will deterministically be split into one photon in each arm after the $50-50$ BS as a result of the Hong-Ou-Mandel effect \cite{Hong1987}. The maximum of the associated probability $\left|\sqrt{3}i\cos^2\theta\sin\theta\right|^2$ is obtained for $\sin^2\theta = \frac{1}{3}$ and thus for a BS with splitting ratio $2:1$.\\
In this configuration, a factor of $\frac{4}{9}$ therefore accounts for probabilistic splitting of the triplet photons so that the triplet coincidence detection rate becomes
\begin{equation}
    R_{\text{triplet}}^{\text{coinc.}} = \frac{4}{9}\eta_q^3R_{\text{triplet}}^{\text{L}},
\end{equation}
where $R_{\text{triplet}}^{\text{L}}$ is the number of triplets per second in the linear configuration L ($x$ $\rightarrow$ $x$,$x$,$x$, cf. subsection \ref{sec:mode_overlap}).\\
At the same time, the effective fluorescence photon rates at each detector are equal for D$_1$, D$_2$, D$_3$, $R_{\text{fl}}^{D_i} = 1/6~R_{\text{fl}}$ ($i=1,2,3$). It follows that the CAR in this case is
\begin{equation}
    \text{CAR} = 96\cdot \frac{1}{\tau^2R_{\text{fl}}^3} R_{\text{triplet}}^{\text{L}}.
\end{equation}
\subsubsection{Circular pump polarization}
In case of circular pump polarization, $2$ of the triplet photons will have the same handedness as the pump whereas one has opposite handedness. Hence, with a combination of a quarter-wave-plate (QWP) and a PBS we can separate the two circular polarizations and add a single $50-50$ BS in the $2$-photon arm (Fig. \ref{fig:detection_schemes} (b): configuration C). The two indistinguishable photons in the first arm are directed onto two different detectors,
\begin{equation}
    R_{\text{triplet}}^{\text{coinc.}} = \eta_q^3R_{\text{triplet}}^{\text{C}},
\end{equation}
but $R_{\text{fl}}^{D_1} = \frac{1}{2}R_{\text{fl}}$, $R_{\text{fl}}^{D_2} = R_{\text{fl}}^{D_3} = \frac{1}{4}R_{\text{fl}}$. Using $R_{\text{triplet}}^{\text{L}}/R_{\text{triplet}}^{\text{C}}\approx1.33$, the CAR in this case is
\begin{equation}
    \text{CAR} = 42.7\cdot \frac{1}{\tau^2R_{\text{fl}}^3} R_{\text{triplet}}^{\text{L}}.
\end{equation}
Considering these scenarios, we can therefore conclude that the optimal configuration for the highest possible CAR is to employ a linearly polarized pump beam combined with detection as in the first detection scheme (Fig. \ref{fig:detection_schemes}(a): configuration L).\\
\subsection{Waist radius fluctuations and detection bandwidth requirements}
The fabrication of sub-micron tapered fibers with a nearly constant waist radius along the full tapered fiber length is challenging. While the perfect phase-matching diameter can be tuned without changing the pump wavelength by placement of the fiber in a pressure tunable gas cell \cite{Cavanna2020,Hammer2018,Hammer2020}, this cannot account for fluctuations of the waist radius along $0\leq z\leq L$. In this section, we consider how these fluctuations affect the expected triplet generation rate. We model these fluctuations as a waist radius $r(z)$ varying along $z$ and following a Gaussian distribution $p(r)\propto e^{-\frac{(r-\mu_r)^2}{2\sigma_r^2}}$ with a mean value $\mu_r$ and standard deviation $\sigma_r$, but do not take fluctuation-induced losses e.g. due to scattering into other modes into account. We will determine which detection bandwidth is optimal in terms of detected coincidence counts and CAR, assuming the first detection scheme (Fig. \ref{fig:detection_schemes} (a)), effectively only collecting photons of the same linear polarization as the pump. This leads to a rate of coincidence counts of $R_{\text{coinc}}=\frac{4}{9}\eta_q^3R_{\text{triplet}}^{\text{L}}$ where $R_{\text{triplet}}^L$ is the integrated triplet rate of configuration L, weighted according to the distribution $p(r)$.\\
\begin{figure}[htpb]
    \centering\includegraphics[width=\textwidth]{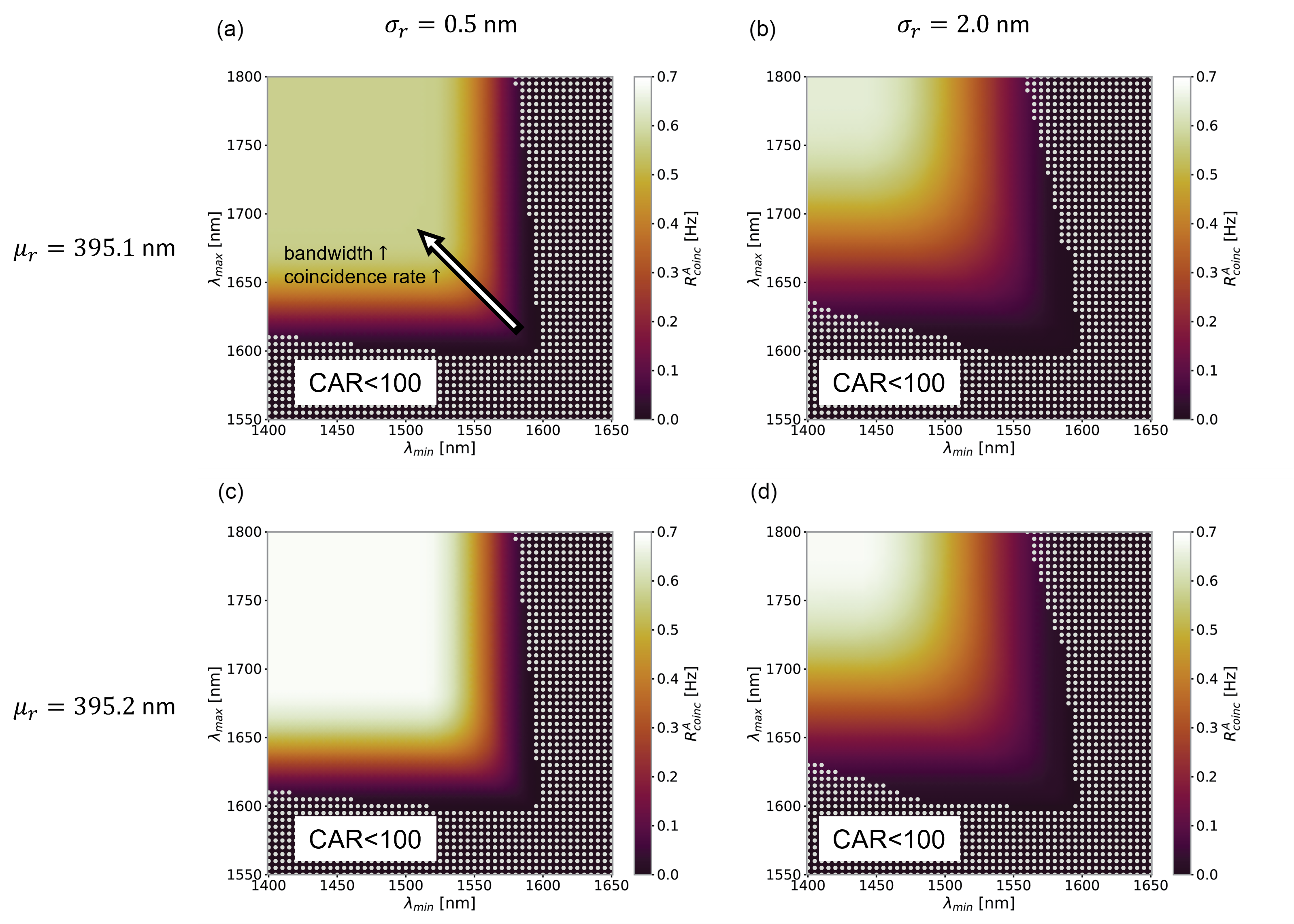}
    \caption{Coincidence rates as obtained for an optical detection bandwidth determined by the detection wavelength limits $\lambda_{min}, \lambda_{max}$ for different Gaussian fiber radius fluctuation distributions centered around $\mu_r$ with standard deviation $\sigma_r$. The dotted area marks the detection settings where the CAR<$100$. \mbox{(a) $\mu_r=395.1$ nm, $\sigma_r=0.5$ nm.} The white arrow illustrates the increasing coincidence rates as the detection bandwidth grows. \mbox{(b) $\mu_r=395.1$ nm, $\sigma_r=2.0$ nm.} \mbox{(c) $\mu_r=395.2$ nm, $\sigma_r=0.5$ nm} \mbox{(d) $\mu_r=395.2$ nm, $\sigma_r=2.0$ nm}}
    \label{fig:coincidence_rates}
\end{figure}
As expected, closer to the perfect phase matching waist radius (Fig. \ref{fig:coincidence_rates} (c) compared to \ref{fig:coincidence_rates} (a)) or with an increased detection bandwidth $B$ (white arrow in Fig. \ref{fig:coincidence_rates} (a)), more coincidences may be detected. For larger waist radius fluctuations ($\sigma=2.0$ nm), the bandwidth necessary to obtain the same coincidence detection rate also increases (Fig. \ref{fig:coincidence_rates} (d) compared to \ref{fig:coincidence_rates} (c)). However, large detection bandwidths are detrimental to the CAR, since the fluorescence count rate scales linearly with the detection bandwidth $B$. Nevertheless, the count rates as calculated here suggest that currently available detection techniques allow for demonstration of the photon triplet state with this setup.

\section{Seeded TOPDC}
We now turn to the case of seeded TOPDC as opposed to spontaneous TOPDC in a tapered fiber. Here, we can take advantage of the fact that the triplet generation rate can be increased by several orders of magnitude with a weak seed beam coupled to the tapered fiber in addition to the pump beam. More specifically, we again assume a pump beam at a wavelength of 532 nm in the higher-order mode HE$_{12}$ and add a seed beam at 1596 nm in the fundamental HE$_{11}$ mode. The interaction Hamiltonian is the same as for the unseeded case,
\begin{equation}
    \hat{H}_I=-\epsilon_0\cdot4\cdot3\cdot2!\cdot\chi^{(3)}_{ijkl}\int_{V_I}d^3\textbf{r}~\hat{E}^{(+)}_{p,i}\hat{E}^{(-)}_{s,j}\hat{E}^{(-)}_{s_1,k}\hat{E}^{(-)}_{s_2,l}~+ \text{h.c.}.
\end{equation}
The seed field is labeled by $\hat{E}_s$, whereas the generated signal photons correspond to the fields $\hat{E}_{s_1},\hat{E}_{s_2}$. The seeded triplet generation rate reads
\begin{equation}
    R_{\text{triplet,seeded}} = \sum_{\xi_{s_1}}\sum_{\xi_{s_2}} \frac{2\pi}{\hbar^2}~\delta(\Delta\omega)~|\langle \hat{H}_I\rangle|^2,
\end{equation}
where
\begin{equation}
    \langle\hat{H}_I\rangle = \langle \alpha(N_p-1),\alpha(N_s+1),1,1|\hat{H}_I|\alpha(N_p),\alpha(N_s),0,0\rangle.
\end{equation}
Initially, both the pump and seed states are coherent states with mean photon numbers $N_p,N_s$ respectively. Since the possible signal modes are closely spaced in $\beta$-space, the sum is transformed into an integral,
\begin{equation}
    R_{\text{triplet,seeded}} =  \sum_{\sigma_{s_1},\sigma_{s_2}} \left(\frac{L}{2\pi}\right)^2\int d \omega_{s_1}\left.\left[\left.\frac{d\beta}{d\omega}\right\vert_{\omega_{s_1}}\left.\frac{d\beta}{d\omega}\right\vert_{\omega_{s_2}}\frac{2\pi}{\hbar^2}~|\langle \hat{H}_I\rangle|^2\right]\right\vert_{\Delta\omega=0}.
\end{equation}
As discussed in section \ref{sec:rate_triplet_generation}, the integration over the interaction volume in the interaction Hamiltonian can be split up into a longitudinal and transversal component. Furthermore using \mbox{$\langle\alpha(N_p-1)|\hat{a}_p|\alpha(N_p)\rangle \approx\sqrt{N_p}$,} \mbox{$\langle\alpha(N_s+1)|\hat{a}^{\dagger}_s|\alpha(N_s)\rangle \approx\sqrt{N_s}$,} \mbox{$\langle1,1|\hat{a}^{\dagger}_{s_1}\hat{a}^{\dagger}_{s_2}|0,0\rangle=1$,}
\begin{equation}
    |\langle\hat{H}_I\rangle|^2 = \frac{36N_pN_s(\chi_{xxxx}^{(3)}/3)^2\epsilon_0^2\hbar^4\omega_p\omega_s\omega_{s_1}\omega_{s_2}}{L^2M_pM_sM_{s_1}M_{s_2}}~\text{sinc}^2\left(\frac{\Delta\beta L}{2}\right)~\mathcal{O}.\\
\end{equation}
Note that the mode overlap $\mathcal{O}$ depends on the signal photon polarizations $\sigma_{s_1},\sigma_{s_2}$. Therefore, the generation rate becomes
\begin{equation}\label{eq:seeded_case_R_triplet}
\begin{split}
    R_{\text{triplet,seeded}} &= \frac{2L^2P_pP_sn_pn_s(\chi_{xxxx}^{(3)})^2\epsilon_0^2}{\pi c^2 M_pM_s}\\
    &~~~~~~~~~~\times\sum_{\sigma_{s_1},\sigma_{s_2}}\int d \omega_{s_1}\left.\left[\frac{\omega_{s_1}\omega_{s_2}}{M_{s_1}M_{s_2}}\left.\frac{d\beta}{d\omega}\right\vert_{\omega_{s_1}}\left.\frac{d\beta}{d\omega}\right\vert_{\omega_{s_2}}\text{sinc}^2{\left(\frac{\Delta\beta L}{2}\right)}\mathcal{O}\right]\right\vert_{\Delta\omega=0}.
\end{split}
\end{equation}
We obtain a total seeded triplet generation rate of over 450 000 Hz in the case of perfect phase matching, a taper waist length of $4$ cm, $\chi_{xxxx}^{(3)}=2.8\times10^{-22}$ mV, a pump power ($x$-polarization) of $100$ mW and a seed beam ($x$-polarization) of $40$ mW in the higher-order HE$_{12}$ mode. The polarization dependence of the mode overlap is exactly the same as for the unseeded, spontaneous TOPDC case considered above so that polarization settings can be adapted according to the experimental requirements. In terms of detection schemes, we note that not only a coincidence detection scheme can be used for state characterization. However, the details pertaining to these prospective methods are out of the scope of this paper.

\section{Four-Wave-Mixing}
\begin{figure}[htpb]
    \centering\includegraphics[width=1.\textwidth]{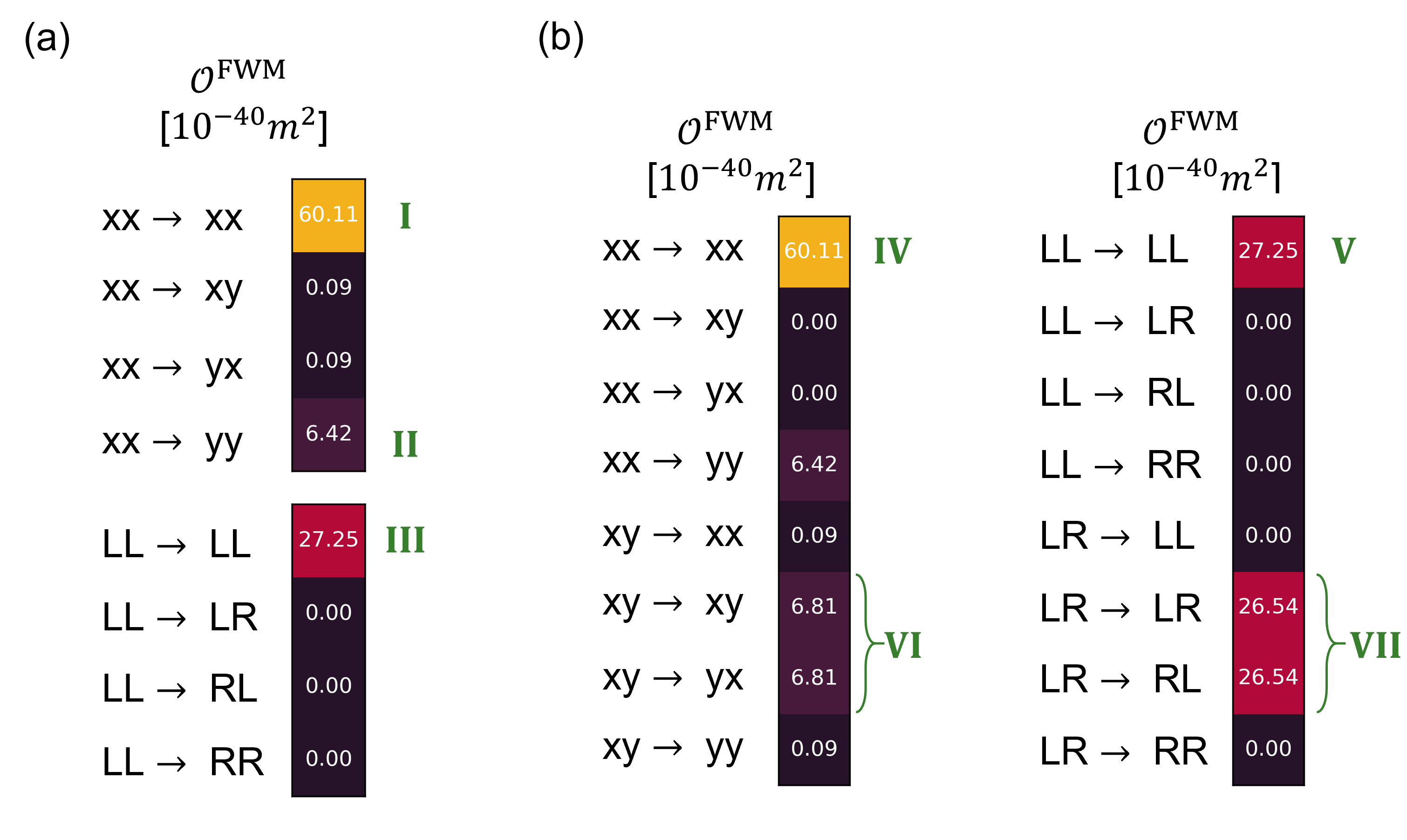}
    \caption{Mode overlap for FWM in the solid-core PCF described above, comparing different pump- and signal-polarization configurations. (a) Linear and circular polarization for a degenerate pump at $722$ nm in the normal dispersion regime. Signal and idler can be generated via FWM at $634$ nm and $836$ nm. While FWM in the $x$,$x$$\rightarrow$$x$,$x$ configuration is most efficient, pair generation is also expected to be non-negligible in both the $x$,$x$$\rightarrow$$y$,$y$ and L,L$\rightarrow$L,L configuration. Note that for circular polarizations, only angular momentum conserving FWM processes are allowed. (b) Linear or circular pump polarization for two non-degenerate pump beams at $634$ nm and $836$ nm. Signal and idler photon pairs of degenerate wavelength may be generated at $722$ nm. While the $x$,$x$$\rightarrow$$x$,$x$ configuration exhibits the highest mode overlap and therefore nonlinear gain, photon pairs can also be expected in the $x$,$x$$\rightarrow$$y$,$y$, the $x$,$y$$\rightarrow$$x$,$y$, the L,L$\rightarrow$L,L and the L,R$\rightarrow$L,R configurations.}
    \label{fig:FWM_mode_overlap}
\end{figure}

Besides an application to TOPDC, a general form of quantized fiber modes of different polarizations lends itself to an analysis of FWM. Polarization effects of FWM or ultrashort pulse propagation in optical fibers have already been theoretically explored for multiple decades \cite{Berkhoer1970,Lin2004,GarayPalmett2007,McKinstrie2004,Poletti2008}. A crucial result was the adherence to angular momentum conservation \cite{Berkhoer1970,GarayPalmett2007,Lin2007,Poletti2008}, which restricts the possible pump and signal/idler configurations in FWM. Accurately addressing signal degradation due to Raman scattering in optical fibers, photon pair polarization correlations have been theoretically investigated in \cite{Lin2007}. Here, we employ the approach above to gain further quantitative insight into the pure FWM efficiency.\\
Just as TOPDC, FWM is mediated by the third-order nonlinear susceptibility of the interaction medium. The interaction Hamiltonian can be written as
\begin{equation}\label{eq:interaction_hamiltonian_fwm}
    \hat{H}_I=-\epsilon_0\cdot4!\cdot\chi^{(3)}_{ijkl}\int_{V_I}d^3\textbf{r}~\hat{E}^{(+)}_{p_1,i}\hat{E}^{(+)}_{p_2,j}\hat{E}^{(-)}_{s_1,k}\hat{E}^{(-)}_{s_2,l}~+ \text{h.c.},
\end{equation}
with $p_1=p_2=p$ in the degenerate FWM case. Because of the frequency dependence, $\chi^{(3)}\equiv\chi^{(3)}(\omega_{p_1}, \omega_{p_2}; \omega_{s_1}, \omega_{s_2})$  is generally different from the tensor we previously considered in the context of triplet generation but obeys the same relations between its components in isotropic media. To examine polarization effects in the quasi-CW case, here we need to take the FWM mode overlap into account,
\begin{equation}
\begin{split}
    \mathcal{O}^{\text{FWM}} &= ~\left|\hat{\chi}^{(3)}_{ijkl}\int_{A_I}dxdy~\mathcal{I}^{\text{FWM}}_{ijkl}(x,y)\right|^2,\\
    \mathcal{I}^{\text{FWM}}_{ijkl}(x,y) &= ~e_{p_1,i}(x,y)e_{p_2,j}(x,y)e_{s_1,k}^*(x,y)e_{s_2,l}^*(x,y).
\end{split}
\end{equation}
In the low-gain regime, the FWM efficiency is proportional to the mode overlap. Just as for TOPDC, there is negligible signal generation as long as the phase-matching condition is not fulfilled. However, due to the usually large peak intensities used in experiments, this phase-matching condition also contains an additional contribution due to the intensity-dependent refractive index change (AC Kerr-effect), $2\beta_{\text{pump}}=\beta_{\text{signal}}+\beta_{\text{idler}}+2\gamma P$. It can be realized in the form of intramodal phase matching, for example in solid-core PCF \cite{Wadsworth2004,Wong2005}. Solid-core PCF have been shown to provide a bright source for generation of photon pairs \cite{Fulconis2005}. Here, we consider a zero-dispersion-wavelength (ZDW) shifted fiber that is modeled using COMSOL Multiphysics \textsuperscript{\textregistered} software with an inner silica diameter of $1.5$ $\mu$m, surrounded by air holes in a honeycomb pattern providing an air filling fraction of $70\%$ as described in \cite{Wong2005}. This fiber was experimentally studied in both the normal and anomalous dispersion regime but in the context of tunable optical parametric generation.\\
The obtained ZDW for the simulated fiber is $725$ nm. In the normal dispersion regime close to the ZDW, e.g. for a pump wavelength of $722$ nm and a power of $1.0$ W, the gain is maximal for widely separated sidebands at $634.85$ nm and $836.88$ nm, following from the phase matching condition.\\
Figure \ref{fig:FWM_mode_overlap} (a) shows a relative comparison of the obtained mode overlaps for different pump, signal and idler polarizations, directly related to the sideband gain. The electrical field distributions obtained from the simulations have been normalized to equal total intensities, permitting a relative quantitative comparison. 
Indeed, for configurations that are not angular momentum conserving, there is none or only neglectable gain. We want to emphasize this result since it holds despite the considered PCF structure not perfectly obeying cylindrical symmetry.\\
In accordance with \cite{GarayPalmett2007} and contrary to the case of TOPDC, the total gain is not independent of pump polarization but significantly lower for a circularly polarized pump (III). The $x$,$x$$\rightarrow$$y$,$y$ configuration mode overlap (II) is almost $90\%$ smaller than the one we obtain in the $x$,$x$$\rightarrow$$x$,$x$ configuration (I). In practice, as mentioned, a main degradation source of the obtained correlation of generated pairs is spontaneous Raman scattering (spRS) \cite{Lin2007}. The degenerate pump setting requires co-polarization of the pump fields, so that spRS can only be sufficiently suppressed if the signal and idler are cross-polarized to the pump (case II; not for cases I or III). While configuration I is therefore favorable in the case of widely separated sidebands at hand, configuration II is beneficial for small detuning, a result matching the detailed calculations involving spRS in \cite{Lin2007}.\\
On the other hand, the same PCF structure could potentially provide a source of frequency-degenerate photons at $722$ nm when pumping at $634$ nm and $836$ nm, assuming a power of $500$ mW at each of the pump frequencies. Then, the pump beams need not be co-polarized. Again, the pair generation rate is linearly related to the mode overlap pertaining to a specific set of polarizations. Fig. \ref{fig:FWM_mode_overlap} (b) summarizes the results we obtain in this case. In case of large frequency detuning and low spRS-induced correlation degradation, the linear polarization configuration IV proves advantageous over the circular polarization configuration V. However, as pointed out in \cite{Lin2007}, if the pump and signal detuning is within the Raman gain bandwidth, the circular polarization configuration VII is superior over VI because of a higher mode overlap and superior over IV due to suppression of spRS-induced noise.\\
Highlighting that configurations II and VII only benefit the correlation signal if spRS plays a significant role, linear co-polarized signal, idler and pump beams (I, IV) always provide an enhanced mode overlap $\mathcal{O}^{\text{FWM}}$. Owing to advances in the manufacturing of noble gas-filled hollow-core fibers \cite{Azhar2013} with high nonlinearity \cite{Balciunas2015} and findings of large sideband separation in ZDW-shifted PCF, these polarization configurations are expected to be more suitable for a realization of enhanced fiber-based pair generation sources, spectroscopy or imaging with undetected photons \cite{Huang2023} or single-photon frequency shifting \cite{Dong2017}.
\clearpage
\section{Conclusion}
Starting with a general expression for quantized electromagnetic fields in optical fibers, we have investigated the role of polarization in TOPDC and FWM using a full tensor description of these third-order nonlinear processes. We found that the relative efficiencies are primarily governed by the mode overlap, which is strongly dependent on polarization states. We could predict polarization correlations of triplet photons in TOPDC as well as signal and idler photons in FWM. The analysis has enabled us to optimize the detection scheme for photon triplet states regarding to the signal-to-noise ratio and calculate concrete expected triplet rates. Consequences for ideal experimental design of fiber-based FWM processes are obtained by a quantitative analysis of the mode overlap in FWM involving different polarization configurations.

\section*{Funding}
The authors acknowledge the Deutsche Forschungsgemeinshaft (DFG, German Research Foundation - \mbox{JO 1090/3-2}) and the International Max Planck Research School Physics of Light (IMPRS-PL) for financial support.

\section*{Acknowledgments}
The authors thank Maria V. Chekhova for constructive scientific discussions about this work.

\section*{Disclosures}
The authors declare no conflicts of interest.

\section*{Data availability}
The authors confirm that the theoretical derivations supporting the findings of this study are available within the article and its supplementary materials.
The data and numerical analysis is available from the corresponding authors, upon reasonable request.

\end{document}


\maketitle

\section{Classical field energy}
The classical electromagnetic field energy in an optical fiber (or general nonmagnetic dielectrics) reads
\begin{equation}
    \mathcal{E} = \frac{1}{2}\int_V d^3\textbf{r} \left[\frac{d}{d\omega}(\epsilon\epsilon_0\omega)\textbf{E}^2+\frac{1}{\mu_0}\textbf{B}^2\right],
\end{equation}
where the integration runs over the full quantization volume which extends over the fiber length $L$ in $z$-direction and over the complete infinite plane in $x$- and $y$- direction,
\begin{equation}
    V = (-\infty,\infty)\times(-\infty,\infty)\times[0,L].
\end{equation}
Integration by parts yields
\begin{equation}
\begin{split}
    \int_V d^3\textbf{r}\textbf{B}^2 &= \int_V d^3\textbf{r}(\nabla\times\textbf{A})\cdot(\nabla\times\textbf{A})\\
    &= -\oint_{\partial V}((\nabla\times\textbf{A})\times\textbf{A})\cdot dS + \int_V d^3\textbf{r}(\nabla\times(\nabla\times\textbf{A}))\cdot\textbf{A}.
\end{split}
\end{equation}
The surface terms becomes zero with appropriately chosen boundary conditions. The second term can be evaluated using
\begin{equation}
\begin{split}
    \nabla\times(\nabla\times\textbf{A}) &= \alpha(\nabla\times(\nabla\times \textbf{f})) + \alpha^*(\nabla\times(\nabla\times \textbf{f}^*))\\
    &= \alpha\frac{\epsilon\omega^2}{c^2}\textbf{f} + \alpha^*\frac{\epsilon\omega^2}{c^2}\textbf{f}^*\\
    &= \omega^2\mu_0\epsilon_0\epsilon(\alpha\textbf{f} + \alpha^*\textbf{f}^*),
\end{split}
\end{equation}
therefore,
\begin{equation}
\begin{split}
    [\nabla\times(\nabla\times\textbf{A})]\cdot\textbf{A} &= \omega^2\mu_0\epsilon_0\epsilon(\alpha\textbf{f} + \alpha^*\textbf{f}^*) \cdot (\alpha\textbf{f}+\alpha^*\textbf{f}^*)\\
    &= \omega^2\mu_0\epsilon_0\epsilon(\alpha^2\textbf{f}^2+\alpha^{*2}\textbf{f}^{*2} + 2\alpha\alpha^*\textbf{f}\cdot\textbf{f}^*).
\end{split}
\end{equation}
In an optical fiber, the mode profiles $\textbf{f}^{(*)}(\textbf{r})$ vary as $e^{\pm i\beta z}$, hence, upon integration over the full fiber length $L$,
\begin{equation}
\begin{split}
    \int_V d^3\textbf{r} \textbf{B}^2 &=\int_V d^3\textbf{r}(\nabla\times(\nabla\times\textbf{A}))\cdot\textbf{A}\\
     &= \int_V d^3\textbf{r} 2\omega^2\mu_0\epsilon_0\epsilon(\textbf{r})\alpha\alpha^*\textbf{f}\cdot\textbf{f}^*.
\end{split}
\end{equation}
Furthermore,
\begin{equation}
\begin{split}
    \int_V d^3\textbf{r} \textbf{E}^2 &=-\int_V d^3\textbf{r} \omega^2(\alpha\textbf{f}-\alpha^*\textbf{f}^*)^2\\
     &= -\int_V d^3\textbf{r} \omega^2(\alpha^2\textbf{f}^2+\alpha^{*2}\textbf{f}^{*2}-2\alpha\alpha^*\textbf{f}\cdot\textbf{f}^*)\\
     &= \int_V d^3\textbf{r} 2\omega^2\alpha\alpha^*\textbf{f}\cdot\textbf{f}^*,
\end{split}
\end{equation}
again using the $e^{\pm i\beta z}$ dependence of $\textbf{f}$. Thus, the complete expression for the classical field energy is
\begin{equation}
\begin{split}
    \mathcal{E} &= \frac{1}{2}\int_V d^3\textbf{r} \left[\frac{d}{d\omega}(\epsilon\epsilon_0\omega)\textbf{E}^2+\frac{1}{\mu_0}\textbf{B}^2\right]\\
    &= \frac{1}{2}\int_V d^3\textbf{r}\left[\frac{d}{d\omega}(\epsilon\epsilon_0\omega)(2\omega^2\alpha\alpha^*\textbf{f}\cdot\textbf{f}^*)+\frac{1}{\mu_0}(2\omega^2\mu_0\epsilon_0\epsilon(\textbf{r})\alpha\alpha^*\textbf{f}\cdot\textbf{f}^*)\right]\\
    &= \omega^2|\alpha^2|\int_V d^3\textbf{r}\left[\left(\frac{d}{d\omega}(\epsilon\epsilon_0\omega)+\epsilon\epsilon_0\right)|\textbf{f}|^2\right]\\
    &= 2\omega^2|\alpha^2|\int_V d^3\textbf{r}\left[\frac{\frac{d}{d\omega}(\epsilon\omega)+\epsilon}{2}\epsilon_0|\textbf{f}|^2\right].
\end{split}
\end{equation}
\section{Triplet generation Interaction Hamiltonian}
\begin{equation}
\begin{split}
    \hat{H}_I &= -24\epsilon_0\chi^{(3)}_{ijkl}\int_{V_I}d^3\textbf{r} \left(i\sqrt{\frac{\hbar\omega_p}{2}}a_pf_{p,i}(\textbf{r})\right)\\
    &~~~~~~~~~\left(-i\sqrt{\frac{\hbar\omega_{s_1}}{2}}a^{\dagger}_{s_1}f^*_{s_1,j}(\textbf{r})\right)\left(-i\sqrt{\frac{\hbar\omega_{s_2}}{2}}a^{\dagger}_{s_2}f^*_{s_2,k}(\textbf{r})\right)\left(-i\sqrt{\frac{\hbar\omega_{s_3}}{2}}a^{\dagger}_{s_3}f^*_{s_3,l}(\textbf{r})\right) + h.c.\\
    &= 24\chi^{(3)}_{ijkl}\epsilon_0~\sqrt{\frac{\hbar\omega_p}{2}}\sqrt{\frac{\hbar\omega_{s_1}}{2}}\sqrt{\frac{\hbar\omega_{s_2}}{2}}\sqrt{\frac{\hbar\omega_{s_3}}{2}}~a_pa^{\dagger}_{s_1}a^{\dagger}_{s_2}a^{\dagger}_{s_3}~\frac{1}{L^2}\frac{1}{\sqrt{M_pM_{s_1}M_{s_2}M_{s_3}}}\\
    &~~~~~~~~~\int_{L_I}dze^{i(\beta_p-\beta_{s_1}-\beta_{s_2}-\beta_{s_3})z}~\int_{A_I}dxdy~e_{p,i}(x,y)e_{s_1,j}^*(x,y)e_{s_2,k}^*(x,y)e_{s_3,l}^*(x,y) + h.c.\\
    &= 24\chi^{(3)}_{ijkl}\epsilon_0~\sqrt{\frac{\hbar\omega_p}{2}}\sqrt{\frac{\hbar\omega_{s_1}}{2}}\sqrt{\frac{\hbar\omega_{s_2}}{2}}\sqrt{\frac{\hbar\omega_{s_3}}{2}}~a_pa^{\dagger}_{s_1}a^{\dagger}_{s_2}a^{\dagger}_{s_3}~\frac{1}{L^2}\frac{1}{\sqrt{M_pM_{s_1}M_{s_2}M_{s_3}}}\\
    &~~~~~~~~~\int_{L_I}dze^{i(\beta_p-\beta_{s_1}-\beta_{s_2}-\beta_{s_3})z}~\int_{A_I}dxdy~\mathcal{I}_{ijkl}(x,y) + h.c.\\
\end{split}
\end{equation}
\clearpage
\section{$\langle\alpha(N_p-1)|a_p|\alpha(N_p)\rangle$ in the limit of large photon numbers}
\noindent
In order to evaluate the triplet generation rate, the approximation
\begin{equation}
\begin{split}
    \langle\alpha(N_p-1)|a_p|\alpha(N_p)\rangle &= \left(e^{-\frac{N_p-1}{2}}\sum_{m=0}^{\infty}\frac{(N_p-1)^{\frac{m}{2}}}{(m!)^{\frac{1}{2}}}\langle m|_p\right)a_p\left(e^{-\frac{N_p}{2}}\sum_{n=0}^{\infty}\frac{N_p^{\frac{n}{2}}}{(n!)^{\frac{1}{2}}}|n\rangle_p\right)\\
    &=e^{-\frac{N_p-1}{2}}e^{-\frac{N_p}{2}}\left(\sum_{m=0}^{\infty}\frac{(N_p-1)^{\frac{m}{2}}}{(m!)^{\frac{1}{2}}}\langle m|_p\right)\left(\sum_{n=1}^{\infty}\frac{N_p^{\frac{n}{2}}}{(n!)^{\frac{1}{2}}}\sqrt{n}~|n-1\rangle_p\right)\\
    &=e^{-\frac{N_p-1}{2}}e^{-\frac{N_p}{2}}\left(\sum_{m=0}^{\infty}\frac{(N_p-1)^{\frac{m}{2}}}{(m!)^{\frac{1}{2}}}\langle m|_p\right)\left(\sum_{n=0}^{\infty}\frac{N_p^{\frac{n+1}{2}}}{(n!)^{\frac{1}{2}}}|n\rangle_p\right)\\
    &=\sqrt{N_p}~ e^{-\frac{N_p-1}{2}}e^{-\frac{N_p}{2}}\left(\sum_{m=0}^{\infty}\frac{(N_p-1)^{\frac{m}{2}}}{(m!)^{\frac{1}{2}}}\right)\left(\sum_{n=0}^{\infty}\frac{N_p^{\frac{n}{2}}}{(n!)^{\frac{1}{2}}}\right)\delta_{mn}\\
    &=\sqrt{N_p}~ e^{-\frac{N_p-1}{2}}e^{-\frac{N_p}{2}}\left(\sum_{n=0}^{\infty}\frac{(\sqrt{(N_p-1)N_p})^n}{n!}\right)\\
    &=\sqrt{N_p}~ e^{-\frac{N_p-1}{2}}e^{-\frac{N_p}{2}}~e^{\sqrt{(N_p-1)N_p}}\\
    &\approx\sqrt{N_p}
\end{split}
\end{equation}
holds in the limit of large photon numbers since \begin{equation}
    \lim_{N_p\rightarrow\infty}\frac{N_p-1}{2}+\frac{N_p}{2}-\sqrt{(N_p-1)N_p}=0.
\end{equation}
\clearpage
\section{Mode overlap for triplet generation in a tapered optical fiber: LCP-LCP-LCP triplet configuration}
Here we show the mode overlap components in cylindrical coordinates for an LCP pump and triplet generation in the LCP-LCP-LCP configuration.
\begin{figure}[htpb]
    \centering\includegraphics[width=0.7\textwidth]{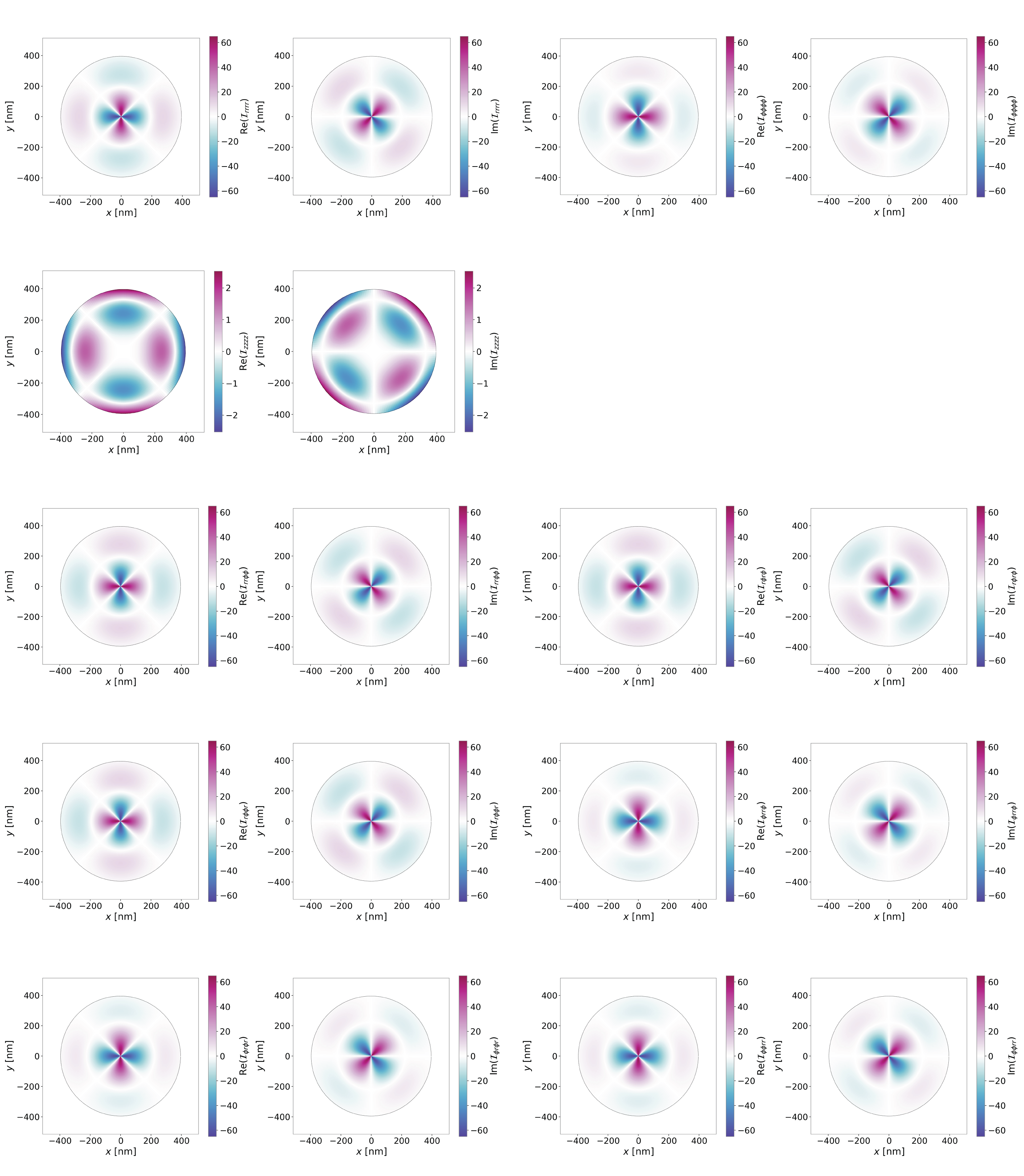}
    \caption{Real and imaginary parts of all components of the mode overlap integrand $\mathcal{I}_{ijkl}$ for the LCP $\rightarrow$ 3$\times$LCP configuration (part 1).}
    \label{fig:si_I_LLLL_1}
\end{figure}
\begin{figure}[htpb]
    \centering\includegraphics[width=0.7\textwidth]{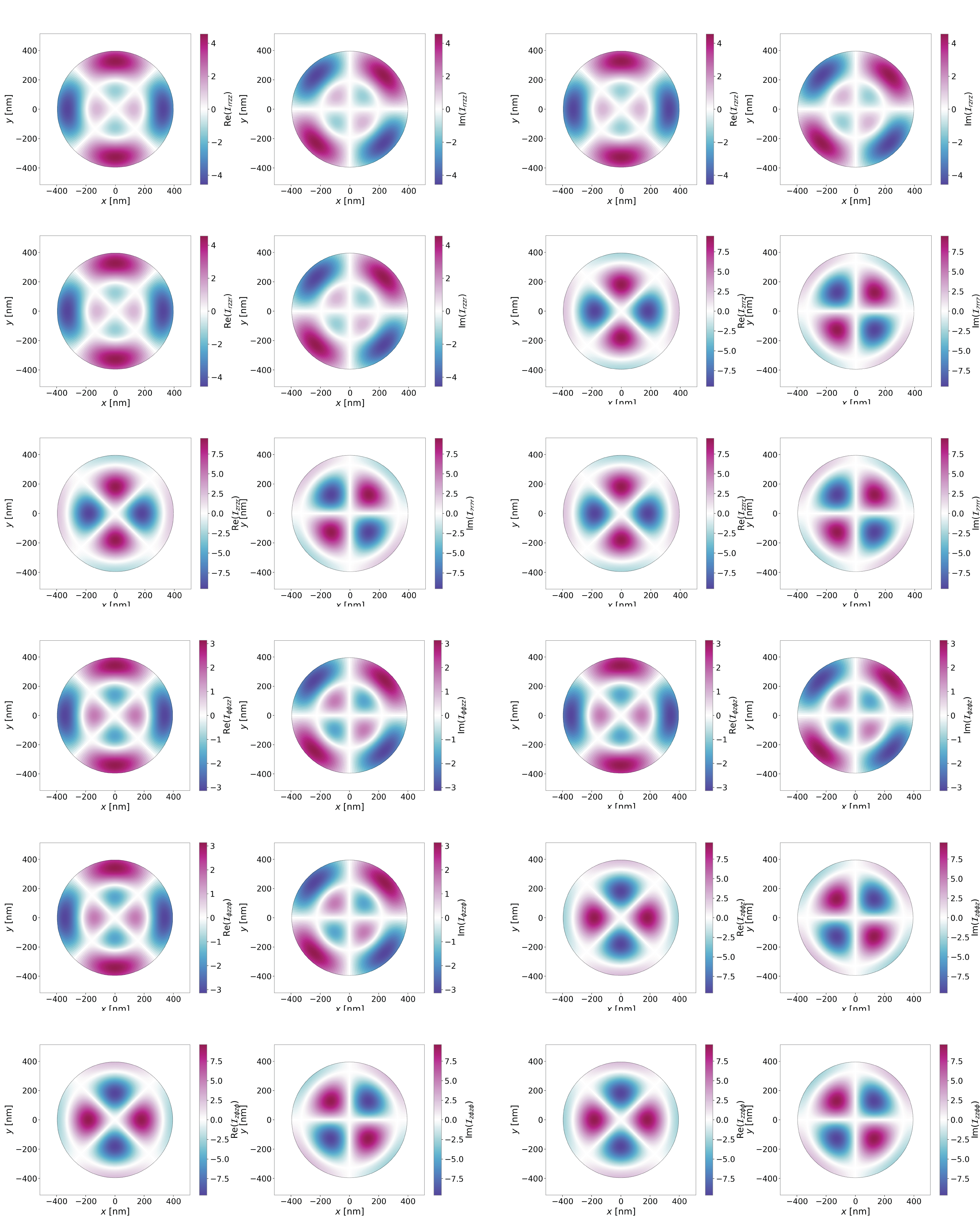}
    \caption{Real and imaginary parts of all components of the mode overlap integrand $\mathcal{I}_{ijkl}$ for the LCP $\rightarrow$ 3$\times$LCP configuration (part 2).}
    \label{fig:si_I_LLLL_2}
\end{figure}
\clearpage
\section{Mode overlap for triplet generation in a tapered optical fiber: LCP-RCP-RCP, RCP-RCP-RCP triplet configuration}
\begin{figure}[htpb]
    \centering\includegraphics[width=0.7\textwidth]{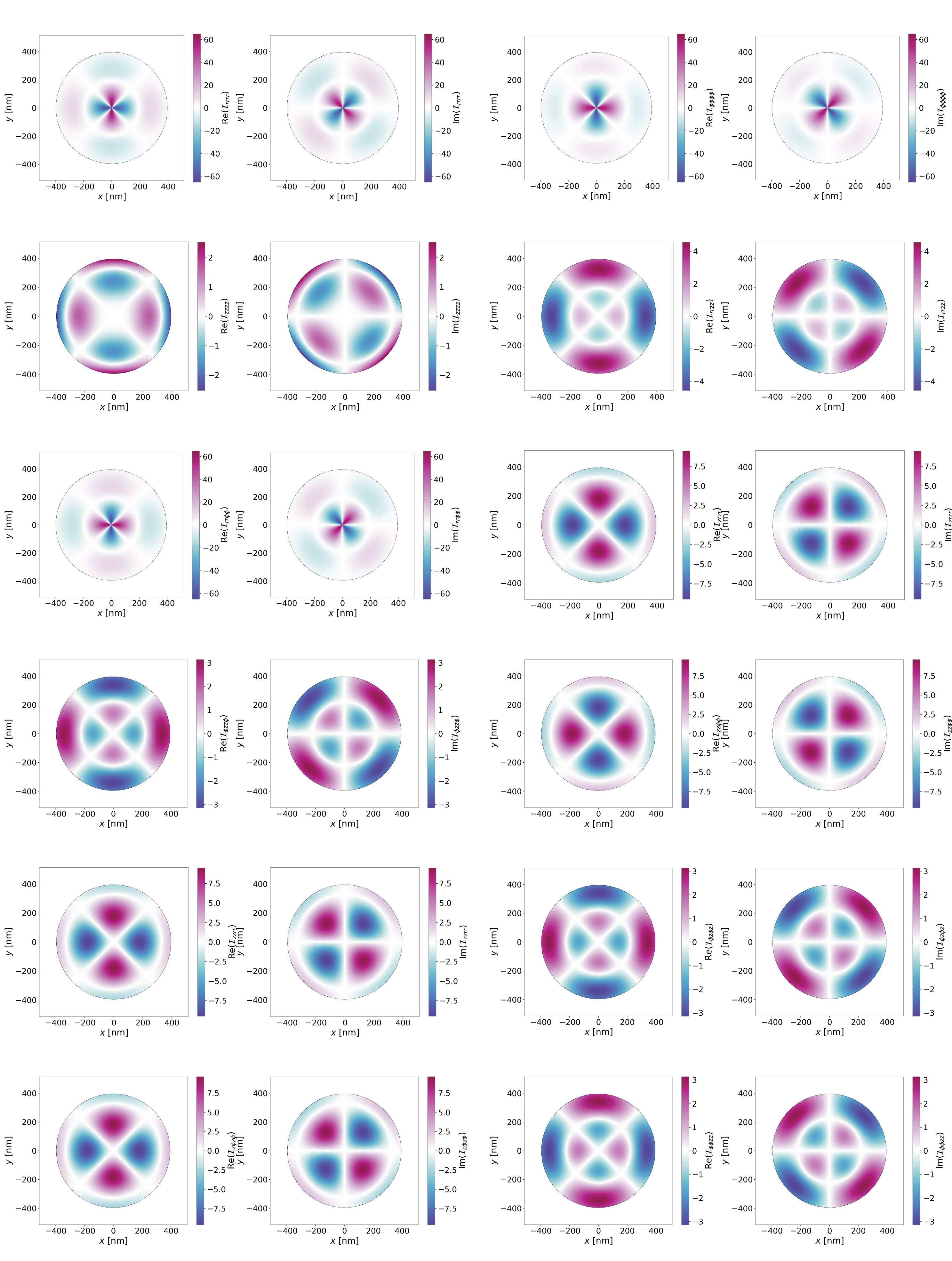}
    \caption{Real and imaginary parts of components of the mode overlap integrand $\mathcal{I}_{ijkl}$ for the LCP $\rightarrow$ 1$\times$LCP,2$\times$RCP configuration.}
    \label{fig:si_I_LLRR}
\end{figure}
\begin{figure}[htpb]
    \centering\includegraphics[width=0.7\textwidth]{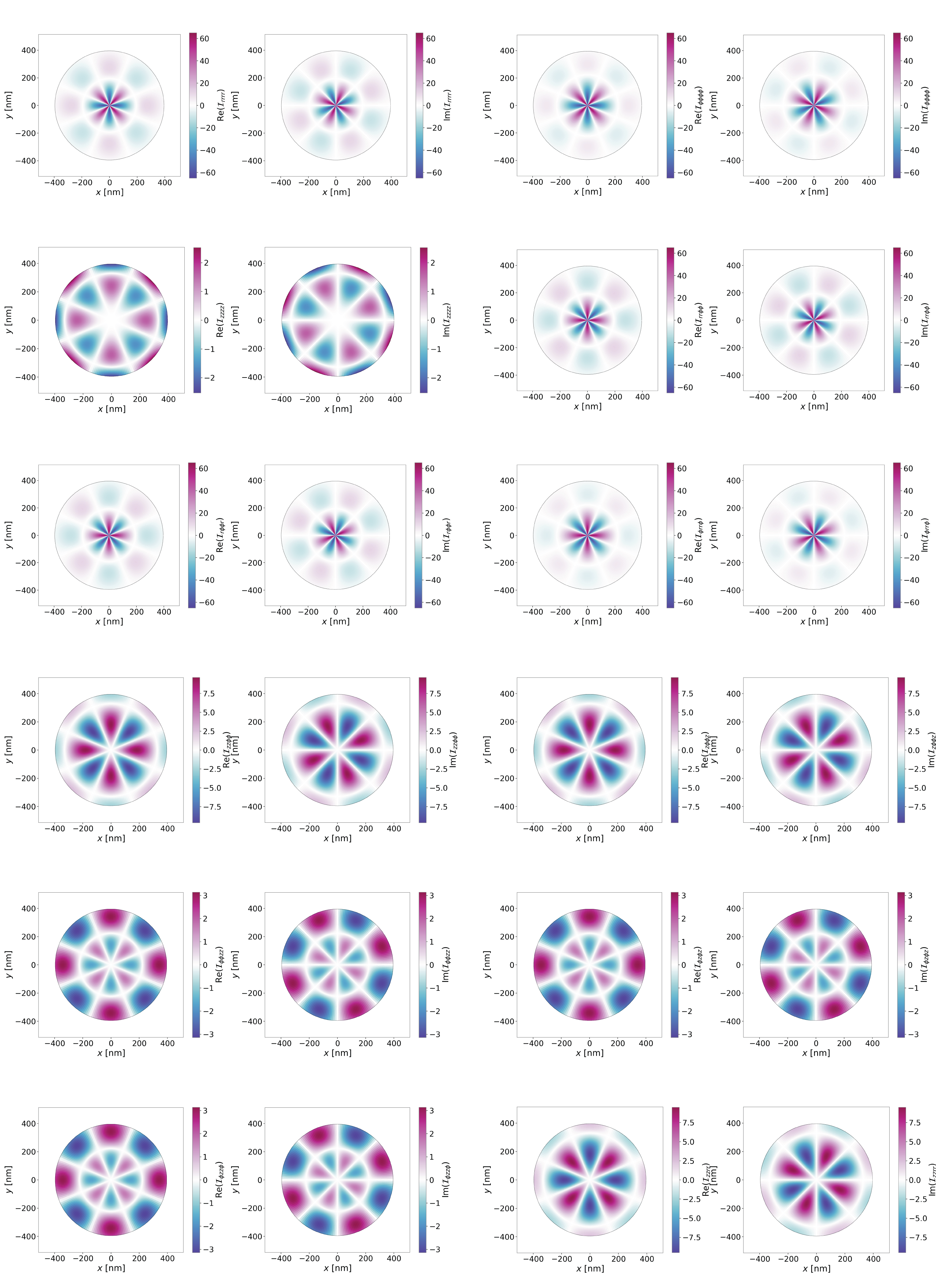}
    \caption{Real and imaginary parts of components of the mode overlap integrand $\mathcal{I}_{ijkl}$ for the LCP $\rightarrow$ 3$\times$RCP configuration.}
    \label{fig:si_I_LRRRR}
\end{figure}